\documentstyle[eqsecnum,aps,twocolumn]{revtex}
\input epsf

\def\be{\begin{equation}}
\def\ee{\end{equation}}
\def\bea{\begin{eqnarray}}
\def\eea{\end{eqnarray}}

\def\cmm2{{\,\rm cm^{-2}}}
\def\cm2{{\,{\rm cm}^2}}
\def\cmm3{{\,{\rm cm}^{-3}}}
\def\gcmm3{{\,{\rm g\,cm^{-3}}}}

\def\ga{\mathrel{\mathpalette\fun >}}
\def\fun#1#2{\lower3.6pt\vbox{\baselineskip0pt\lineskip.9pt
  \ialign{$\mathsurround=0pt#1\hfil##\hfil$\crcr#2\crcr\sim\crcr}}}
\def \etal{{\it et al.} }
\def \etai{{\eta_{\rm i}}}
\def \etaf{{\eta_{\rm f}}}
\def \etaobs{{\eta_{\rm obs}}}
\def \hatbfn{{\hat{\bf n}}}
\def \bfx {{\bf x}}
\def \bfk {{\bf k}}
\def \bfxobs{{\bfx_{\rm obs}}}
\def \rmI{{\rm I}}
\def \rmS{{\rm S}}
\def \rmV{{\rm V}}
\def \rmT{{\rm T}}
\def \s{\scriptscriptstyle}
\def \Ylm{{Y_{\s(l,m)}}}
\def \Deltalm{{\Delta_{\s(l,m)}}}

\begin{document}
\twocolumn[\hsize\textwidth\columnwidth\hsize\csname @twocolumnfalse\endcsname
\preprint{HEP/123-qed}
\title{CMB Anisotropy Induced by Cosmic Strings\\ 
on Angular Scales $\ga 15'$}
\author{B. Allen$^1$, R.~R. Caldwell$^2$, S. Dodelson$^3$, L. Knox$^4$, \\
E.~P.~S. Shellard$^5$, and A. Stebbins$^3$}
\address{$^1$ Dept. of Physics, University of Wisconsin - Milwaukee, P.O. Box
413, Milwaukee, Wisconsin 53201, USA}
\address{$^2$ Dept. of Physics and Astronomy, University of Pennsylvania,
Philadelphia, Pennsylvania 19106, USA}
\address{$^3$ NASA/Fermilab Astrophysics Center, P. O. Box 500, Batavia, Illinois 60510, USA}
\address{$^4$ Canadian Institute for Theoretical Astrophysics, Toronto, Ontario M5S 3H8, CANADA}
\address{$^5$ University of Cambridge, D.A.M.T.P. Silver Street, Cambridge CB3 9EW, UK}
\date{\today}
\maketitle

\begin{abstract}
We have computed an estimate of the angular power spectrum of the Cosmic
Microwave Background (CMB) induced by cosmic strings on angular scales
$\ga15'$, using a numerical simulation of a cosmic string network; and
decomposed this pattern into scalar, vector, and tensor parts.  We find no
evidence for strong acoustic oscillations in the scalar anisotropy but rather a
broad peak.  The anisotropies from vector modes dominate except on very
small angular scales while the tensor anisotropies are sub-dominant on all
angular scales. The anisotropies generated after recombination are even
more important than in adiabatic models.  We expect that these qualitative
features are robust to the varying of cosmological parameters, a study which
has not yet been done.
\vskip 0.2in \end{abstract}


]


The spectrum of CMB anisotropy on the angular scales somewhat smaller than
that subtended by the horizon at last scattering provides a powerful probe
of the nature of inhomogeneities and matter in our universe\cite{Scott}. On
these scales hydrodynamical effects can leave characteristic signatures of
adiabatic or isocurvature perturbations\cite{HUSUG}, active or passive
perturbations\cite{Active}, an open or closed universe, or even a high or low
Hubble constant.  Cosmic strings are topological defects which may have formed
in the very early Universe and may be responsible for the formation of large
scale structure observed in the Universe today \cite{strings}.  While cosmic
string induced perturbations are clearly both isocurvature and active the
interplay between these two properties allows a range of possible behavior for
the degree scale anisotropy\cite{ActiveConfusion}.  One method for determining
the signature for cosmic strings is via simulation, i.e. by numerically
evolving a network of cosmic strings in a simulated universe and computing the
pattern of anisotropy they would produce.  We present the results of just such
a calculation in this {\it Letter}.  While we have not explored the full range
of cosmological parameters, or included all the effects on network dynamics,
our results are suggestive of what is and is not important for small scale
anisotropies from the cosmic string model.

\section{Methodology}

Cosmic strings only interact gravitationally with the rest of the matter and
produce only small metric perturbations.  Thus we may ignore back-reaction of
the metric perturbations on the strings and solve for the evolution of the
string network in an unperturbed cosmology.  This ``stiff-source''
approximation allows us to do the string simulation and later compute the
perturbations to the matter.  It also allows us to reuse the the same string
simulation to compute the perturbations in cosmologies with different matter
content.  Here we have reused one of the simulations that was was used in Allen
\etal \cite{Allen96}.  While that simulation took the size of the simulation
box to be twice the present day horizon, we may use the assumption of
self-similar evolution of the string network (``scaling'') to rescale the box
to a smaller size while simultaneously decreasing both the starting and ending
time of the simulation.    The equations of motion of the strings will still
be satisfied so long as the cosmological expansion remains a power law, which
in this case is $a\propto t^{2\over3}$, corresponding to a flat FRW matter
dominated universe.  These simulations should not be used for epochs too close
to matter-radiation equality, which means that if we want to study the effects
close to recombination we should really only consider a large Hubble constant
which puts recombination long after matter-radiation equality.  Here we use
$H_0=80$, but even with this Hubble constant the effects of the
matter-radiation transition are liable to be significant, and below we discuss
what these effects are liable to be. Throughout we  
take the 
baryon fraction $\Omega_b = .02$ to be consistent with the
predictions of nucleosynthesis.

	The cosmic string simulation is in a cubical box whose comoving size we
denote by $L$. This simulation starts when the comoving horizon ($\equiv$
conformal time) given by $\etai=0.05L$ and ends at $\etaf=0.5 L$.
During this period the universe has grown by a factor of 100.  The final time
was chosen to the maximal value for which one is guaranteed that the 
network properties will not be affected by the periodic boundary conditions
put on the string network.  The statistical properties of CMB anisotropy
generated in this box will also not be affected by the periodicity so long as
the anisotropies are measured at conformal time $\etaobs\le\etaf$.  In
order to use this simulation to compute small-scale anisotropies produced near
recombination one must make the measurement of the anisotropies long after the 
end of the string simulation.  One may do this by rescaling the simulation to
end just after recombination, but continuing to propagate the brightness
pattern unperturbed through the periodic box until today.  This {\it
free-streaming} procedure will lead to a brightness pattern which is
quasi-periodic on the sky, the size of the periodic patches being the angle
subtended by length $L$ when placed at the distance of the
surface-of-last-scattering.  We expect that the statistical properties of the
pattern in each of these patches will be nearly the same as for a similar sized
patch in an infinite non-periodic universe. 

Since the metric perturbations are small we may use the linearized Einstein
equations for the metric and the linearized Boltzmann equation for the
anisotropies.  This makes the equations for  the temperature anisotropy 
linear in the stress-energy of the strings, $\Theta_{\mu\nu}$.  The dynamics of
the strings is described by non-linear equations, and leads to the non-Gaussian
distribution for $\Theta_{\mu\nu}$, but the {\it response} of the photons to
the strings is linear.  The solution of the linear equations may be written as
a homogeneous and inhomogeneous part
\begin{eqnarray}
&&\hskip-15pt{\Delta T\over T}(\hatbfn,\bfxobs,\etaobs)=
                                       \left({\Delta T\over T}\right)^\rmI+ \cr
&&\hskip-5pt
\int d\eta'\int d^3\bfx' D^{\mu\nu}(\hatbfn,\bfx,\bfx',\eta,\eta')\,
                          \Theta_{\mu\nu}(\bfx'\hatbfn')
\label{Pattern}
\end{eqnarray}
where $({\Delta T\over T})^\rmI$  depends on the initial condition for the
cosmological fluids (CDM, baryons, photons, and neutrinos) which is often
referred to as {\it compensation}.  We have compensated the strings with 
adiabatic perturbations in the other matter  so as to make the total density
initially uniform, just as in Ref.~\cite{Allen96}.  Clearly this is not exactly
the correct prescription except on superhorizon scales. Nevertheless we find
that $({\Delta T\over T})^\rmI$ is small in comparison with the 2nd term 
(this is consistent with the results of Ref. \cite{Caldwell97}). We
think this is likely to remain true with different prescriptions for
compensation.

	The way we have used Eq.~(\ref{Pattern}) is somewhat different than has
typically been done in the past.  Rather than concentrating on a few $\bfxobs$
we will effectively compute the entire brightness pattern throughout the
simulation box.  This is a function of the 5-d phase space $(\hatbfn,\bfxobs)$
and is quite a large object.  One can reduce this size by Fourier decomposing
the $\bfx$-dependence and then, for each $\bfk$-mode, expanding the 
$\hatbfn$-dependence in spherical harmonics, $\Ylm$, using a spherical polar
coordinate system with the pole in the direction of $\bfk$. The beauty of this
decomposition is that the $m=0$ terms correspond to scalar modes, the $m=\pm1$
terms to vector modes, and the $m=\pm2$ to tensor modes\cite{AbbottSchaefer}.
Modes with $|m|>2$ are zero since they are not coupled to the gravitational
field (in linear theory) and may be ignored.  From the mode coefficients,
$\Deltalm(\bfk)$, one may construct the volume average of the $C_l$'s 
 measured
by all the observers in the box: 
\begin{equation}
\overline{C_l}=\overline{C_l}^\rmS+\overline{C_l}^\rmV+\overline{C_l}^\rmT
\end{equation}
where
\begin{eqnarray}
\overline{C_l}^\rmS&=&{1\over(2l+1)\,L^3}
               \sum_\bfk|\Delta_{\s(l, 0)}|^2                               \cr
\overline{C_l}^\rmV&=&{1\over(2l+1)\,L^3}
         \left(\sum_\bfk|\Delta_{\s(l,+1)}|^2+|\Delta_{\s(l,-1)}|^2\right)  \cr
\overline{C_l}^\rmT&=&{1\over(2l+1)\,L^3}
         \left(\sum_\bfk|\Delta_{\s(l,+2)}|^2+|\Delta_{\s(l,-2)}|^2\right)\ .
\end{eqnarray}
There are no cross-terms between different terms because all of the modes are
orthogonal. Applying this decomposition to the right-hand-side of
Eq.~(\ref{Pattern}) induces a scalar/vector/tensor decomposition of
$D^{\mu\nu}$ and $\Theta_{\mu\nu}$.  We may perform the decomposition of
$\Theta_{\mu\nu}$ numerically.  The corresponding components of $D^{\mu\nu}$
are solutions of the linearized Einstein-Boltzmann equation which we have
computed numerically using standard techniques\cite{DJK}.

\section{Tests}

Our calculation has three parts: the string simulation, the numerical
calculation of the Green functions, and the computation of the
brightness perturbation, which merges the two preceding parts.  A
number of tests have been made on the simulation
\cite{Allen96,Allen90,Caldwell97} which we do not discuss here.

The Green functions have been tested in two ways. First, the code which
computes the scalar and tensor Green functions \cite{DJK} has been used
to determine the angular power spectrum for a model with adiabatic
perturbations only.  These calculations agree to within a percent with
standard results. Although the vector Green functions cannot be tested
in this way, there is an analytic approximation in terms of spherical
Bessel functions, which holds after recombination. For this case, there
is excellent agreement between the analytic and numerical results.
Similarly, the scalar Green functions have been tested on large scales
by comparing with analytic approximations; again we find excellent
agreement.

To test the full pipeline of merging the Green functions with the
stress energies, we calculated the anisotropies in a matter dominated
universe with one domain wall\cite{SHOBA}. The $C_l$'s in this case can
be calculated analytically. The numerical results are in very good
agreement with the exact results.

The greatest limitation of our ability to accurately compute the small
angle CMB anisotropy is the lack of dynamic range of the cosmic string
simulation. Although the simulation runs over a range of $10$ in
conformal time, this is insufficient to accurately follow the evolution
of modes from well outside to inside the horizon, where they begin to
oscillate. Our results indicate that modes well inside the horizon
make a significant contribution to the anisotropy.

Due to the limited spatial resolution of the cosmic string simulation,
a limited number of Fourier modes of the source stress energy are
available. To determine whether this would prevent us from seeing
``Doppler peaks,'' we used the same number of modes over the same range
of time for a model with adiabatic initial conditions, for which the
exact results are known. Qualitative agreement was obtained, as the
peak structure and location were clearly evident. Hence, we do not
expect that the limited spatial resolution should prevent us from
resolving features in the angular power spectrum.  Finite grid effects
lead to an artificial drop of only $\sim 10\%$ in the amplitude of the
angular power spectrum for the largest $l$-values obtained from a
simulation volume.

Finally, we have used the same simulation to calculate anisotropies
generated at different times. However, we do not expect there to be
strong correlations in either the source stress energy or the
anisotropy on time intervals longer than the run time of the string
simulation. Hence, we do not expect the re-cycling of the numerical
simulation to have a strong effect on our results.

\section{Results}

We obtain the final angular power spectrum, $C_l$, by combining the
results from different rescalings of the results of the cosmic string
numerical simulation.  As discussed earlier, we exploit the scaling
properties of the string network to make up for the finite temporal
dynamic range of the simulation. For multipole moments in the range $2
<  l  < 20$ (large angles) we use one box, covering the redshift range
$0 < z < 100$.  For $100 <  l  < 800$ (small angles) we use two
different boxes corresponding to $700 < z < 17000$ and $60 < z < 700$.

Our results are shown in Figures 1 and 2.
On large angular scales, we see that the dominant source of the
anisotropy is due to vector perturbations. This is an important result,
as it means the COBE normalization of the mass-per-unit-length fixes
the amplitude of the vector, not scalar, spectrum. 

We may use the large angle results as a test of our techniques, by comparing
with the results of Ref. \cite{Allen96}, which used the same simulation.  They
found $l(l+1)C_l \sim 350 (G\mu)^2$ at large angles, so our results are a
factor of $\sim 2.8$ smaller.  However, we are in accord with the shape of the
spectrum and the determination that the majority of the large angle anisotropy
is generated at redshifts $z \lesssim 20$ (as in Fig. 3 of Ref.
\cite{Allen96}). When normalizing to COBE\cite{COBE}, we find 
$\mu_6 \equiv G\mu \times 10^{6} = 1.7$, a factor $\sim\sqrt{2.8}$ higher than
that of Ref. \cite{Allen96}.  This result is comparable with $\mu_6 = 1.5 (\pm
0.5)$ \cite{Bennett92}, $1.7 (\pm 0.7)$ \cite{perivo} and $2$
\cite{coulson}. Thus, our large angle normalization falls in the middle of the
range spanned by other calculations.

On small angular scales Figure 1 shows the contribution to the scalar $C_l$'s
from early times ($z > 700$).  There is a gradual rise in the spectrum from
$l=100$ till a very broad plateau for $l\ga 200$. The increase is less than a
factor of two so may not be significant.

Calculating the contribution from the vector modes and from the late time
contributions to the scalar modes presents a problem.  In each case, the box
size must be chosen very large in order to get to very late times. The smallest
scales ($l \ga 300$) therefore become unreliable. In Figure 2, we show the
results for $100 < l <300$.  The most striking feature here is that both the
vector contribution and the late time scalar contribution are larger than
the early time scalars. The net result is that the mild acoustic peak in the
scalar spectrum becomes hidden.  The total spectrum therefore is quite
flat. (It is unclear if the slight drop in the total is significant.)  While
these qualitative features appear to be robust, we believe that in future work
they will be subject to quantitative changes which will tend to significantly
boost power on small-angular scales.

\section{Discussion}

We have calculated the anisotropies in the CMB and the perturbations to
matter induced by a network of cosmic strings in the matter era.  This
calculation uses {\it all} the components of the stress-energy tensor
and exact Green functions for all modes. The greatest source of
uncertainty in our results is due to the limited spatial resolution and
run time of the numerical simulation.

On large scales, we find that the vector perturbations are very
important for the CMB anisotropy from cosmic strings. Our results for
anisotropies at small scales appear at present to be too low to be
consistent with current measurements.  Future work will determine
whether inclusion of a radiation-matter transition in the cosmic string
simulation, in which the long string density and rms velocity relaxes
from the radiation- to matter-era scaling values, inclusion of a
``wiggly'' equation of state for the cosmic strings, or a lower value
of the Hubble parameter and a larger value of the baryon density can
boost the small-scale power sufficiently.  Even if a primary Doppler
peak were to become a more prominent feature, this work suggests that
secondary Doppler peaks are not anticipated in cosmic string models.
Finally let us mention that qualitatively similar results have also been found
for global defects\cite{PST97}.

\vspace{36pt}

The work of BA was supported by NSF Grants No. PHY91-05935 and No.
PHY95-07740. The work of RRC was supported by DOE Grant No.
DOE-EY-76-C-02-3071. The work of SD and AS was supported in part by DOE
and NASA grant NAGW-2788 at Fermilab.  EPS is partially supported by
PPARC grant GR/H71550. 

\begin{figure}[p]
\caption{Angular Power Spectrum at large and small scales for the
cosmic string simulation. Tensors do not contribute appreciably.
On small scales only the contribution from early times ($z >700$) can
be calculated reliably over the full range of angular scales.}
\label{fig:cls}
\end{figure}

\begin{figure}[p]
\caption{Angular Power Spectrum at small scales for the
cosmic string simulation. The three different contributions
are: the early time scalars; late time scalars; and vectors. These
add incoherently to give the total spectrum.}
\label{fig:cltotal}
\end{figure}

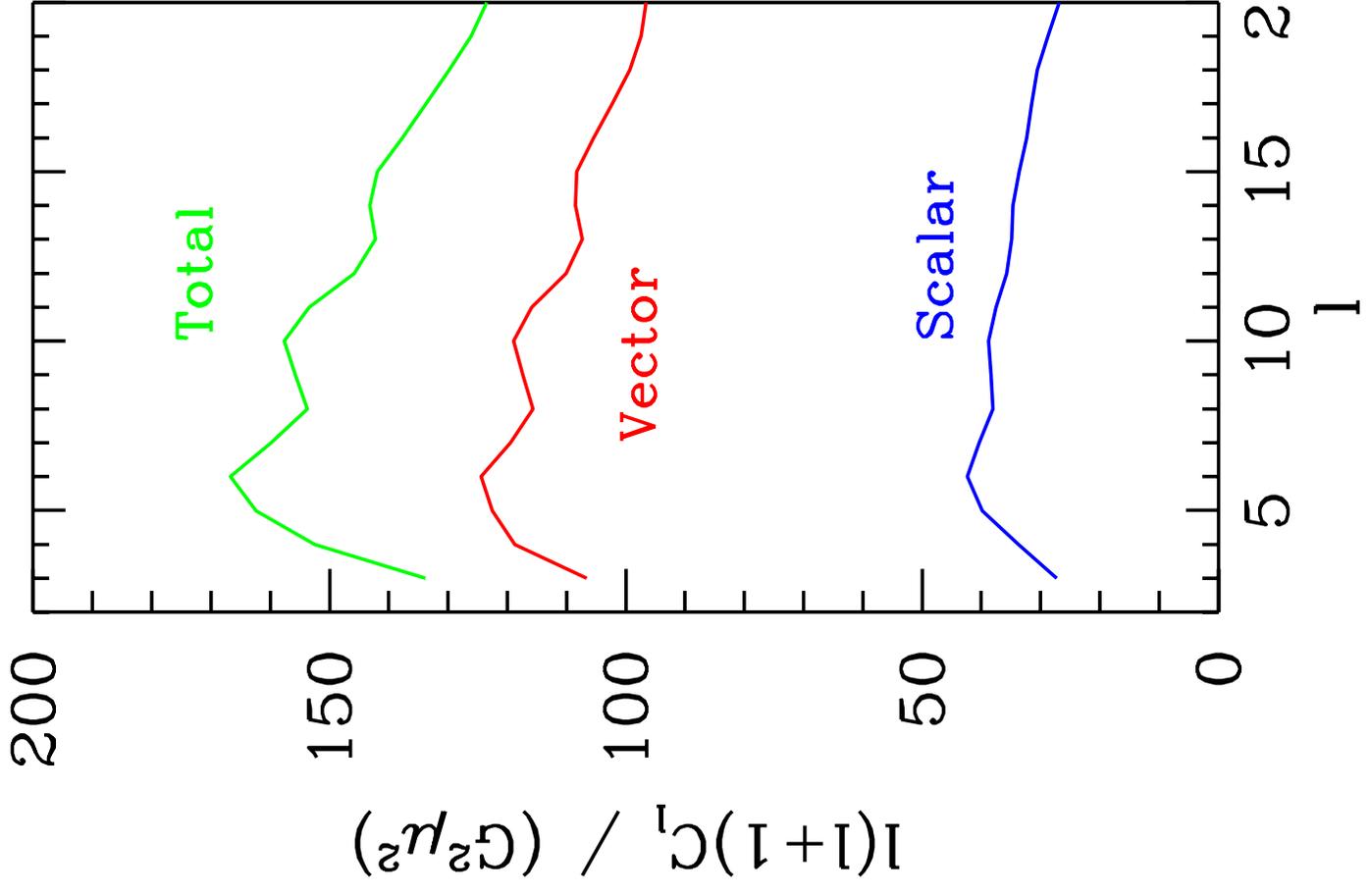
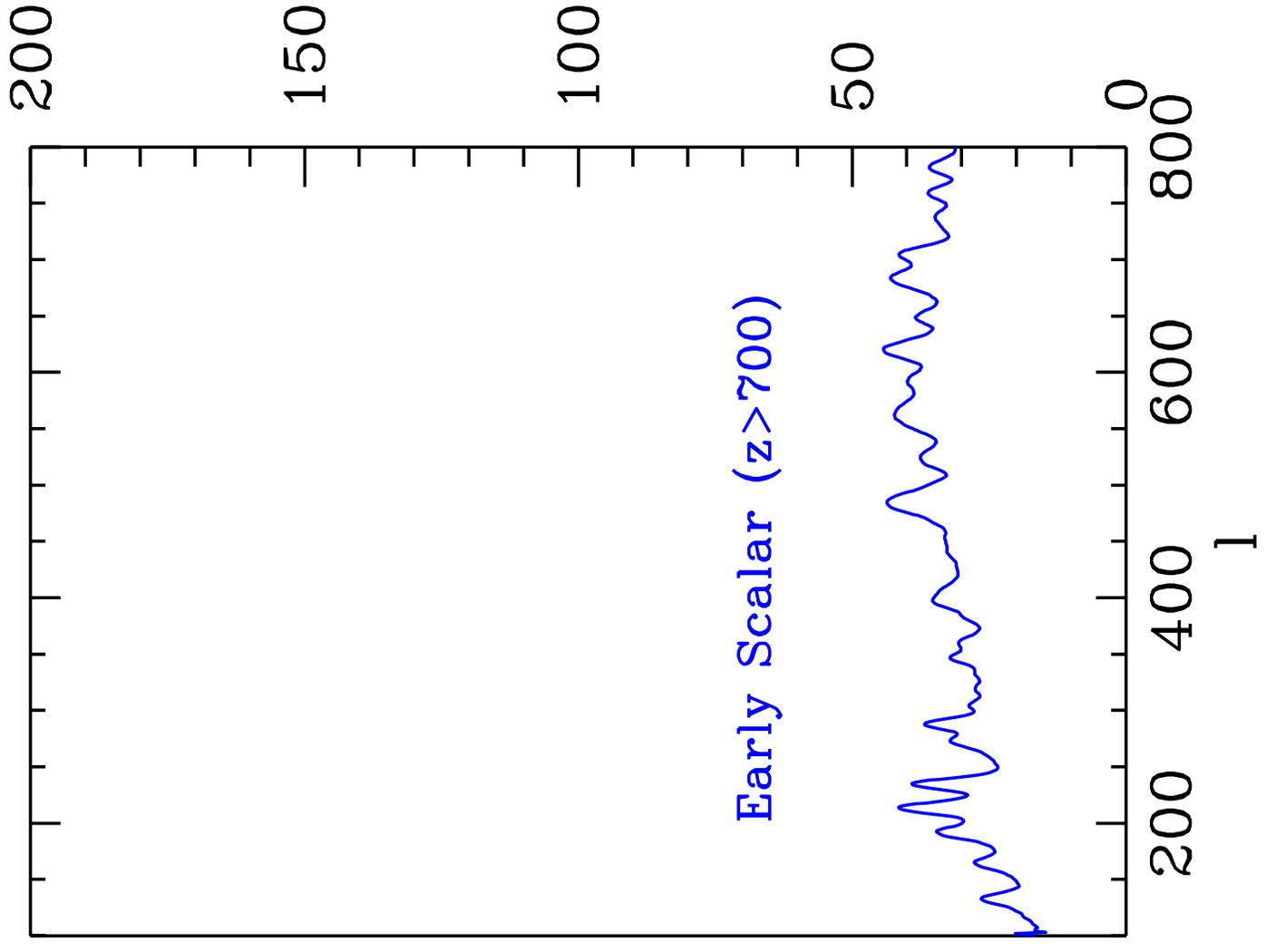
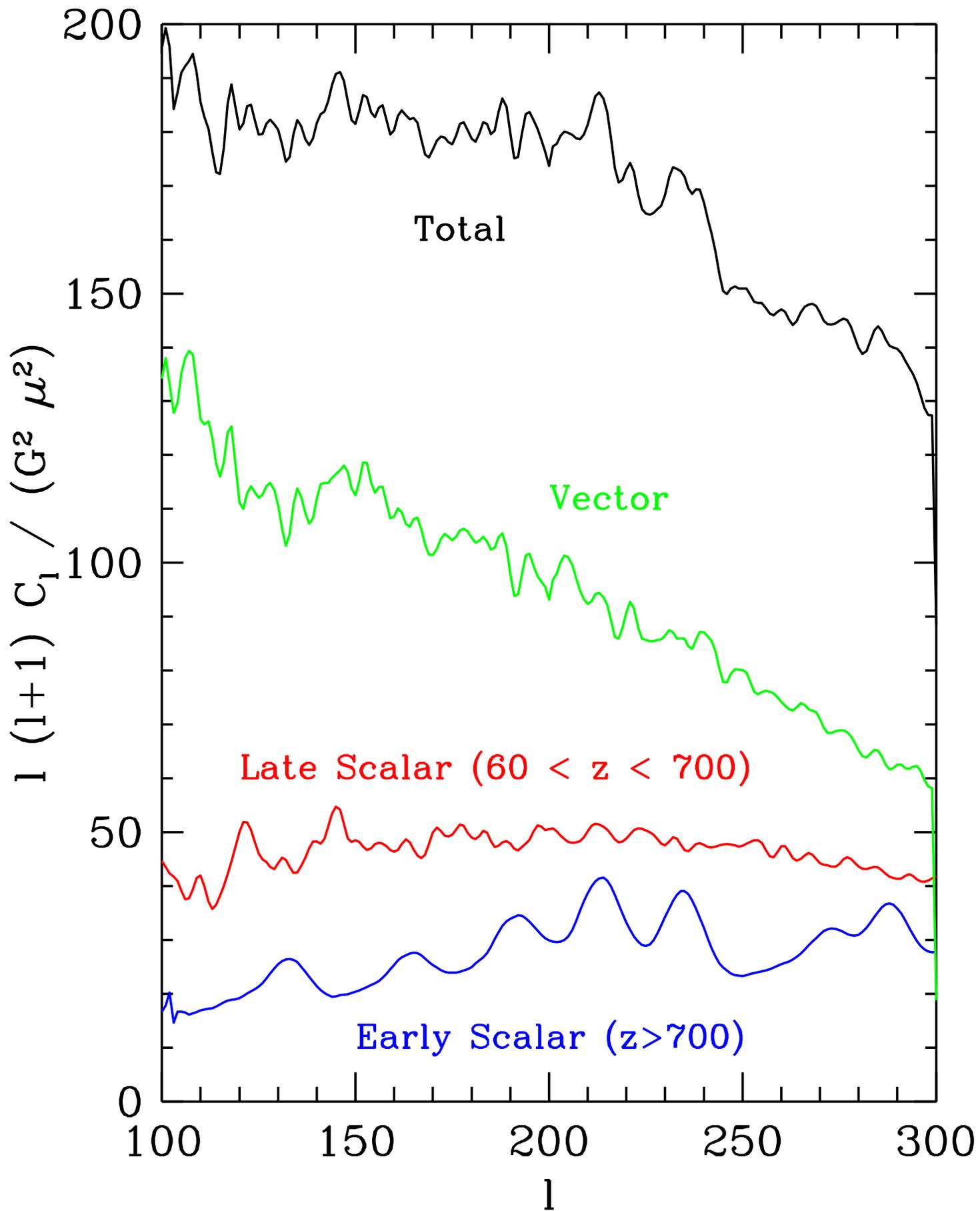
\end{document}